\documentstyle[12pt]{article}

\newcommand{\bge}{\begin{equation}}
\newcommand{\ee}{\end{equation}}
\newcommand{\mixten}[3]{{#1}^{#2}_{\phantom{{#2}} #3}}

\newcommand{\startappendix}{
\renewcommand{\thesection}{\Alph{section}}}

\begin{document}
\begin{titlepage}
\begin{center}

January 15, 1993    \hfill    PUPT-1376  \hskip .8truein      LBL-33458
\hskip .5truein    UCB-PTH-93/02\\

\vskip .5in

{{\Large    \bf  Two   Dimensional  QCD is a  String  Theory}}
\footnote{This work was supported in part by the Director, Office of Energy
Research, Office of High Energy and Nuclear Physics, Division of High
Energy Physics of the U.S. Department of Energy under Contract
DE-AC03-76SF00098 and in part by the National Science Foundation under
grant PHY90-21984.}

\vskip .15in

David J. Gross\footnote{\tt gross@pupphy.princeton.edu}\\[.15in]
{\small {\em
\begin{tabular}{ccc}
\makebox[2.3in]{Joseph Henry Laboratories} & & Theoretical Physics Group\\
Princeton University & and & \makebox[2.3in]{Lawrence Berkeley Laboratory} \\
Princeton, New Jersey 08544 & & 1 Cyclotron Road \\
& &Berkeley, California 94720
\end{tabular}}}

\vskip .15in
Washington Taylor IV\footnote{\tt wati@physics.berkeley.edu}\\[.15in]

{\small {\em
%
%
\begin{tabular}{ccc}
\makebox[2.3in]{Department of Physics} & & \makebox[2.3in]{Theoretical
Physics Group}\\
University of California & and & Lawrence Berkeley Laboratory \\
Berkeley, California 94720 & & 1 Cyclotron Road \\
& &Berkeley, California 94720
\end{tabular}}}
\end{center}

\vskip .15in

\begin{abstract}
The partition function of two dimensional QCD
on a Riemann surface of area $A$ is expanded as a power series in $1/N$
and $A$.  It is shown that the coefficients of this expansion are  precisely
determined  by a sum over maps from a two dimensional surface onto
the two dimensional  target space.  Thus two dimensional QCD has a simple
interpretation as a closed string theory.
\end{abstract}
\end{titlepage}
\renewcommand{\thepage}{\roman{page}}
\setcounter{page}{2}
\mbox{ }

\vskip 1in

\begin{center}
{\bf Disclaimer}
\end{center}

\vskip .2in

\begin{scriptsize}
\begin{quotation}
This document was prepared as an account of work sponsored by the United
States Government.  Neither the United States Government nor any agency
thereof, nor The Regents of the University of California, nor any of their
employees, makes any warranty, express or implied, or assumes any legal
liability or responsibility for the accuracy, completeness, or usefulness
of any information, apparatus, product, or process disclosed, or represents
that its use would not infringe privately owned rights.  Reference herein
to any specific commercial products process, or service by its trade name,
trademark, manufacturer, or otherwise, does not necessarily constitute or
imply its endorsement, recommendation, or favoring by the United States
Government or any agency thereof, or The Regents of the University of
California.  The views and opinions of authors expressed herein do not
necessarily state or reflect those of the United States Government or any
agency thereof of The Regents of the University of California and shall
not be used for advertising or product endorsement purposes.
\end{quotation}
\end{scriptsize}

\vskip 2in

\begin{center}
\begin{small}
{\it Lawrence Berkeley Laboratory is an equal opportunity employer.}
\end{small}
\end{center}

\newpage
\renewcommand{\thepage}{\arabic{page}}
\setcounter{page}{1}

\section {Introduction}
\setcounter{equation}{0}

\baselineskip 18.5pt

It has long been an outstanding problem to find an exact description
of gauge theories in terms of a theory of strings. Recently one of us
(DG) has explored this possibility for two dimensional QCD and has
conjectured that the free energy of a gauge theory with gauge group
$SU(N)$ or $U(N)$ on a 2-dimensional manifold ${\cal M}$ can be
identified with the partition function of a closed, orientable, string
theory with target space ${\cal M}$ \cite{gross}.

The partition function of a pure gauge theory
can be easily calculated on an arbitrary orientable 2-dimensional manifold
${\cal M} $ of genus $G$ and area $A$, (essentially due to the fact that the
action is invariant under area preserving diffeomorphisms), and is given by
\cite{migdal}
 \begin{eqnarray}
{\cal Z}_{\cal M} & = &  \int[{\cal D} A^\mu]
e^{- {1\over 4 {\tilde g}^2} \int_{\cal M} d^2 x\sqrt{g}
 Tr F^{\mu \nu}  F_{\mu \nu}} \\
 & = &  {\cal Z}  (G,\lambda A, N)  = \sum_{R}^{} (\dim R)^{2 - 2G}
  e^{-\frac{\lambda A}{2 N}C_2(R)}, \label{eq:partition}
\end{eqnarray}
where  the sum is taken over all irreducible representations of the gauge
group,
with $\dim R$ and $C_2(R)$ being the dimension and quadratic
Casimir of the
representation $R$. ($\lambda$ is related to the gauge coupling
$\tilde{g}$ by $\lambda = \tilde{g}^2 N$.)
The conjecture states that the free energy
 $ {\cal W}  (G,\lambda A, N) \equiv  \ln {\cal Z}  (G,\lambda A,N)$,
 is equal to the partition function of some string theory with target
space ${\cal M} $, where the string coupling is identified with $1/N$,
and the string tension is identified with $\lambda$.

We do not have a precise formulation of the string action or functional
integral. Instead we shall relate the QCD free energy to a specific sum over
maps. This can be taken as a definition of the string theory--however  we would
still like a  path integral  formulation, particularly in order  to
draw lessons about
string theory in higher dimensions.
Some  things are clear from the structure of the gauge theory. The string
theory must have the symmetries of QCD$_2$--thus it must be invariant under
area preserving diffeomorphisms. This is a feature of the Nambu (and even the
Polyakov) string action.  Since the free energy is an expansion in powers of
$e^{-\lambda A}$  the desired string theory action   must be  proportional to
the area of the map,  like the Nambu action.  However, unlike the Nambu action
it would appear that folds
are totally suppressed. There are two reasons for this. First, as we shall show
below, we can account for the terms in the $1/N$ expansion without invoking
maps with folds.  In other words, the area dependence of the free energy is
totally accounted for by factors of $e^{-{n\lambda A \over 2}}$
arising from maps that
cover the target space precisely $ n$ times, and by powers of the area that
arise from summing over positions of the branchpoints and collapsed handles of
the maps. Second, there is no term in the $1/N$ expansion that behaves as
$e^{-0\lambda A}$, which would correspond to maps with winding number zero.
This is important since in string theory such maps  describe the propagation of
ordinary stringy particles. Even in a two dimensional string theory we would
expect at least one particle--corresponding to the center of mass of the
string. This is the {\em  tachyon} that appears in non-critical two dimensional
string theory with zero mass. However pure two dimensional Yang-Mills theory
contains no particles. If we forbid all folds then we forbid all maps with zero
winding number, since all such maps contain folds. Then   the particles can  be
absent.
Forbidding folds does not render the theory topological, since the
term in the string theory action which suppresses folds need not be
invariant under arbitrary diffeomorphisms of the target
space.\footnote{  A suggestion for modifying the usual Nambu-Goto
string to suppress folds was made by Minahan \cite{Minahan};
however the  term that he  suggested is not invariant under area preserving
diffeomorphisms. } The explicit area-dependence of the theory also
keeps it from being a purely topological theory.

In \cite{gross}  the above conjecture was explored by expanding  the
free energy in
powers of $1/N$--the purported string coupling constant,
\begin{equation}
{\cal W}  (G,\lambda A, N) = \sum_{g = 0}^{\infty} N^{2 - 2g}
f_{g}^{G}(\lambda A).
\label{eq:freeeng}
\end{equation}
It was shown that the coefficients $f_{g}^{G}(\lambda A)$ have precisely the
structure expected if they are given by a sum over maps of  a genus $g$
manifold, ${\cal M}_g$, onto a genus $G$ manifold ${\cal M}_G$, with an action
that is simply the exponential of the area,
\begin{equation}
f_{g}^{G}(\lambda A) = \sum_{n}\sum_{i} \omega_{g,G}^{n,i} \;
e^{- {n \lambda A\over 2}}(\lambda A)^{i}.
\label{eq:expansion}
\end{equation}
The sum over $n$ was interpreted as a sum over maps of winding number
$n$, that
cover the target space $n$ times, and the powers  of the area in
(\ref{eq:expansion}) were interpreted as coming from the contribution of
branchpoints and collapsed handles  of these maps.  The main evidence for this
interpretation was the demonstration that the QCD$_2$  expansion satisfies the
bound, $ g-1 \geq n(G-1)$, that holds for continuous maps of ${\cal M}_g$ onto
${\cal M}_G$ with winding number $n$\cite{gross}.

In this paper  we prove that the term in (\ref{eq:expansion}) with the
maximum
powers of the area for given $g ,G$ and $n$, namely    $ \omega_{g,G}^{n,i}$
with
$2 (g - 1) = 2n (G - 1) + i$, is equal to the number of topologically distinct
continuous maps of
 ${\cal M}_g$ onto ${\cal M}_G$ with winding number $n$ and with $i$
branchpoints. More precisely,
$i! \omega_{g,G}^{n,i}$ is given by the sum of a natural
symmetry factor over all
homotopically distinct $n$-fold connected covers of ${\cal M}_G$.  The
factor of
$i!$  arises because when a genus $g$ surface covers a genus $G$
surface $n$ times, there are precisely $i=2(g - 1) - 2n (G - 1) $ branch
points. Since these branch
points are indistinguishable,  and can be positioned anywhere on the target
space, one obtains a factor of $A^i/i!$.

We will furthermore demonstrate that the remaining terms
$\omega_{g,G}^{n,i}$ with $2 (g - 1) > 2n (G - 1) + i$ can be
interpreted in terms of branchpoints, collapsed handles, and two types
of ``tubes'' connecting sheets of the covering space.    We show that
such objects contribute specific factors, which taken together form a
set of ``Feynman rules'' for the string theory.  (The rules for
collapsed handles and orientation-preserving tubes
were previously suggested by Minahan \cite{Minahan}.)
In the case of
the torus ( $G = 1$), we can interpret all of the coefficients in this
fashion.  For $G \neq 1$, there are some terms of lower order in $N$
for which we do not yet have a geometric interpretation.

In the next section we will discuss the $1/N$ expansion of the gauge
theory defined by (\ref{eq:partition}).  We define an asymptotic
expansion of (\ref{eq:partition}), written in terms of a sum over
Young tableaux.  We show that when written in this fashion, the
partition function factorizes into two separate and identical parts,
or ``chiral sectors'',
coupled by a simple term.  In section 3, we consider
the partition function of a single chiral
sector, and prove that the leading terms in the partition function are
given by the sum of a symmetry factor over all branched covers of
${\cal M} $.  In section 4, we interpret the sectors as corresponding
to covers of opposite orientation, and show that for the torus all of
the remaining terms in the full partition function can be interpreted
in terms of further structures of the covering space which
are localized at a point in ${\cal M} $(collapsed
handles, infinitesimal tubes, etc).  In section 5, we approach
the problem from a more local perspective, and rederive the results of
section 3 by calculating the partition function on a single plaquette
and gluing together plaquettes to form ${\cal M} $.  The results from
this section are also applicable to manifolds with boundary. In section 6, we
briefly discuss how our results generalize to the case of  non-orientable
target spaces.  Finally,
in section 7,  we review our results and discuss further questions. In the
Appendix we derive some properties of the representations of $SU(N)$ needed
for the large $N$ expansion.

\section{$1/N$ Expansion}
\setcounter{equation}{0}

We will now describe the $1/N$ expansion of the partition function
(\ref{eq:partition}) when the gauge group is $SU(N)$. One  might also consider
other groups.  Non semi-simple groups, such as $U(N)$, involve  an extra
coupling and do not appear to have such a nice stringy interpretation. We
shall  therefore  restrict our attention in this paper to  $SU(N)$, although
other simple groups, such as $SO(N)$, which  would correspond to non-orientable
strings, could be analyzed using the same techniques.

The partition function is written as a sum over
representations of the gauge group $SU(N)$.  We will perform an
asymptotic expansion of (\ref{eq:partition}) by first expanding the
contribution from each representation separately as a series in $1/N$,
and then summing over representations.  This is a somewhat ambiguous
procedure, since ``fixing'' a representation as $N \rightarrow \infty$
is not a particularly well-defined process.  As we will see below, the
simplest way of performing this sum only picks out half of the theory.
By carefully defining the set of representations to
sum over, however, we get a theory which seems to contain all the
essential features of the gauge theory at fixed $N$.  We believe that this is
the correct asymptotic expansion of  QCD$_2$, although like all
asymptotic expansions it  does not converge. We would expect non-perturbative
corrections of order exp$ \left(- 1/g_{\rm string} \right)$. These,
however, are not determined by the string perturbation theory (the $1/N$
expansion). To deal with them in string theory one would need the {\em string
field theory }. In our case QCD$_2$  is the string field theory! In fact in
QCD$_2$ there are well defined, order ${e}^{-N}$,  corrections to the
asymptotic expansion, see
\cite{gross}.

\begin{figure}[tbp]
\centering
\begin{picture}(300,200)(-170,-100)
\thicklines

\put(60,- 55){\makebox(0,0){$C_2 (R) = 22 N + 2 - 484/N$}}

\put(-120,80){\line(1,0){175}}
\put(-120,80){\line(0,-1){150}}
\put(-120,55){\line(1,0){175}}
\put(55,80){\line(0,-1){25}}
\put(30,80){\line(0,-1){25}}
\put(-120,30){\line(1,0){125}}
\put(5,80){\line(0,-1){50}}
\put(-20,80){\line(0,-1){50}}
\put(-120,5){\line(1,0){75}}
\put(-120,-20){\line(1,0){75}}
\put(-45,80){\line(0,-1){100}}
\put(-120,-45){\line(1,0){50}}
\put(-120,-70){\line(1,0){50}}
\put(-70,80){\line(0,-1){150}}
\put(-95,80){\line(0,-1){150}}

\put(-140,68){\makebox(0,0){\footnotesize $n_1= 7$}}
\put(-140,43){\makebox(0,0){\footnotesize $n_2= 5$}}
\put(-140,18){\makebox(0,0){\footnotesize $n_3= 3$}}
\put(-140,-7){\makebox(0,0){\footnotesize $n_4= 3$}}
\put(-140,-32){\makebox(0,0){\footnotesize $n_5= 2$}}
\put(-140,-57){\makebox(0,0){\footnotesize $n_6= 2$}}

\put(-108,95){\makebox(0,0){\footnotesize $c_1= 6$}}
\put(-83,110){\makebox(0,0){\footnotesize $c_2= 6$}}
\put(-58,95){\makebox(0,0){\footnotesize $c_3= 4$}}
\put(-33,110){\makebox(0,0){\footnotesize $c_4= 2$}}
\put(-8,95){\makebox(0,0){\footnotesize $c_5= 2$}}
\put(17,110){\makebox(0,0){\footnotesize $c_6= 1$}}
\put(42,95){\makebox(0,0){\footnotesize $c_7= 1$}}

\end{picture}
\caption[x]{\footnotesize A Young tableau for a Representation of $SU(N)$}
\label{f:tableau}
\end{figure}
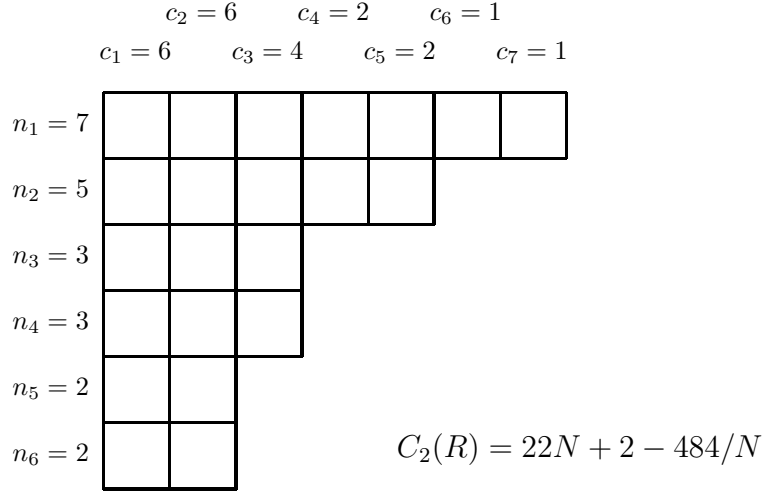

Each representation $R$
is associated with a Young tableau containing some number of
boxes $n$, which are distributed in  rows of length $ {n}_1 \geq {n}_2
\ldots
 \geq  {n}_l  > 0$ and
columns of length $c_1,  \geq c_2 \ldots   \geq  c_k>0 $, where $k =
{n}_1$, and $l = c_1$.  The total
number of boxes is clearly given by $n = \sum_i n_i = \sum_i
{c}_i$  (for an example of a Young tableau, see figure~\ref{f:tableau}).
The quadratic Casimir of the representation $R$ is
given by
\begin{equation}
C_2 (R) = n N + \tilde{C}(R) - \frac{n^2}{N},
\label{eq:casimir}
\end{equation}
where
\begin{equation}
\tilde{C}(R) = \sum_{i} {n}_i^2 - \sum_{i}c_i^2 =\sum_{i}n_i (n_i + 1- 2i) =
\sum_{i}-c_i (c_i + 1 - 2i).
\label{eq:}
\end{equation}
The leading term in a $1/N$ expansion of the dimension of a representation
$R$, whose Young tableau has $n$ boxes, is given by
\begin{equation}
\dim R = \frac{d_{R} N^{n}}{n!} + {\cal O} (N^{n - 1}),
\label{eq:dimension}
\end{equation}
where $d_{R}$ is the dimension of the representation of the symmetric
group $S_{n}$
associated with the Young tableau  $R$.   The $1/N$ correction terms to
(\ref{eq:dimension}) can written in terms of the lengths $n_i$ of the
Young tableau (see the Appendix.) We do not yet have an understanding of the
geometrical
significance of these correction terms, however, and for the most part we will
not
discuss them further in this paper.  Note, though, that for the
torus, which is the physical case of interest, the dimension term does
not appear in the partition function (\ref{eq:partition}).  Thus, we
will be able to completely understand the theory in the case $G = 1$,
without having understood the significance of these correction terms.

\begin{figure}[tbp]
\centering
\begin{picture}(340,230)( - 190,- 130)
\thicklines

\put(100,35){\makebox(0,0){ $R$}}
\put(100,-105){\makebox(0,0){ $S$}}
\put(-70,-120){\makebox(0,0){ $T = \bar{S}R$}}
\put(-140,-100){\dashbox{5}(80,200)}
\put(60,80){\vector(-1,0){50}}
\put(40,-70){\vector(-1,0){90}}
\put(-155,0){\makebox(0,0){$\vdots$}}
\put(-80,0){\makebox(0,0){$\vdots$}}
\put(-100,0){\makebox(0,0){$\vdots$}}
\put(-120,0){\makebox(0,0){$\vdots$}}

\put(-140,100){\line(1,0){140}}
\put(-140,100){\line(0,-1){80}}
\put(-140,80){\line(1,0){140}}
\put(0,100){\line(0,-1){20}}
\put(-140,60){\line(1,0){120}}
\put(-20,100){\line(0,-1){40}}
\put(-40,100){\line(0,-1){40}}
\put(-140,40){\line(1,0){80}}
\put(-140,20){\line(1,0){80}}
\put(-60,100){\line(0,-1){80}}
\put(-80,100){\line(0,-1){85}}
\put(-100,100){\line(0,-1){85}}
\put(-120,100){\line(0,-1){85}}

\put(-140,-20){\line(1,0){80}}
\put(-140,-20){\line(0,-1){60}}
\put(-140,-40){\line(1,0){80}}
\put(-60,-20){\line(0,-1){20}}
\put(-140,-60){\line(1,0){60}}
\put(-80,-15){\line(0,-1){45}}
\put(-140,-80){\line(1,0){40}}
\put(-100,-15){\line(0,-1){65}}
\put(-120,-15){\line(0,-1){65}}

\put(70,100){\line(1,0){60}}
\put(70,100){\line(0,-1){40}}
\put(70,80){\line(1,0){60}}
\put(130,100){\line(0,-1){20}}
\put(70,60){\line(1,0){40}}
\put(110,100){\line(0,-1){40}}
\put(90,100){\line(0,-1){40}}

\put(50,-40){\line(1,0){80}}
\put(50,-40){\line(0,-1){60}}
\put(50,-60){\line(1,0){80}}
\put(130,-40){\line(0,-1){20}}
\put(110,-40){\line(0,-1){20}}
\put(50,-80){\line(1,0){40}}
\put(90,-40){\line(0,-1){40}}
\put(50,-100){\line(1,0){20}}
\put(70,-40){\line(0,-1){60}}

\put(-155,90){\makebox(0,0){\footnotesize 1}}
\put(-155,70){\makebox(0,0){\footnotesize 2}}
\put(-155,50){\makebox(0,0){\footnotesize 3}}
\put(-155,30){\makebox(0,0){\footnotesize 4}}
\put(-155,-30){\makebox(0,0){\footnotesize $N - 3$}}
\put(-155,-50){\makebox(0,0){\footnotesize $N - 2$}}
\put(-155,-70){\makebox(0,0){\footnotesize $N - 1$}}
\put(-155,-90){\makebox(0,0){\footnotesize $N$}}

\end{picture}
\caption[x]{\footnotesize Young Tableau for a Composite Representation.}
\label{f:composite}
\end{figure}
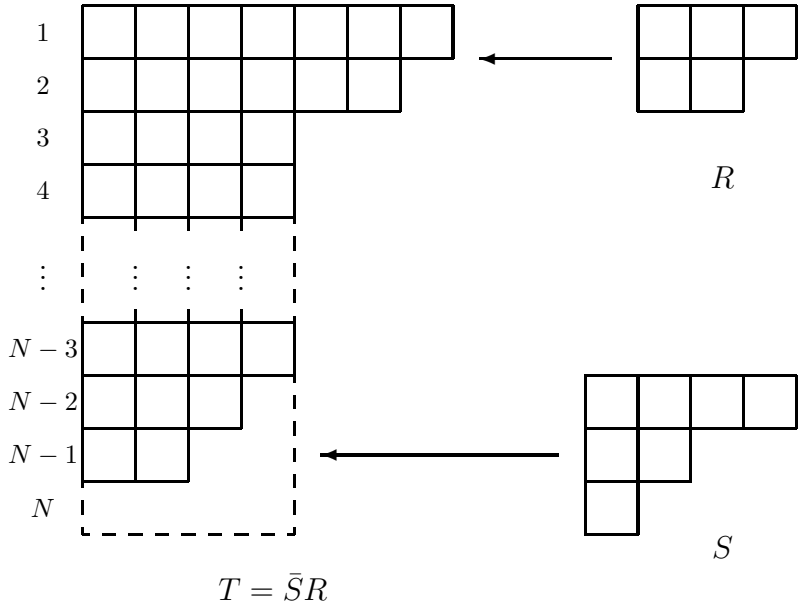

{}From the equation for the quadratic Casimir (\ref{eq:casimir}), one
might expect that to pick out all of the terms in the asymptotic
expansion (\ref{eq:expansion}) which contain terms of the form
$e^{-\frac{n\lambda A}{2}}$, it would suffice to sum over all
representations $R$ in the set $Y_n$ of Young tableaux with $n$ boxes.
This would lead to the partition function
\begin{eqnarray}
\lefteqn{Z (G,\lambda A, N)  = \sum_{n= 0}^{\infty}  \sum_{R \in Y_n}
 (\dim R)^{2 - 2G}  e^{-\frac{\lambda A}{2 N}C_2(R)}}
\\
 & = & \sum_{n}\sum_{R\in Y_n}\left( \frac{n!}{d_R}\right)^{2G - 2}
e^{-\frac{ n \lambda A}{2} } \sum_{i = 0}^{\infty}\left[
\frac{(-\lambda A
\tilde{C}(R))^i}{2^i i!}
N^{n (2 - 2G) - i}+{\cal O}(N^{n (2 - 2G)- i - 1})\right].
\label{eq:partfunct}
\end{eqnarray}
This expansion for the partition function, however, only contains half
of the full theory.  This is because there exists another set of
representations whose quadratic Casimir contains a leading order term
of $n N$.  We will find (\ref{eq:partfunct}) to be a useful object to
study nonetheless, since it contains much of the physics of the full
theory.\footnote{ In \cite{gross}
 these other representations were ignored. This does not, as we shall see,
modify the conclusions of that paper.}
We will now discuss the other representations of interest.  Consider
two representations $R$ and $S$, whose Young tableaux contain $n$ and
$\tilde n$ boxes respectively, with columns
of length $c_i$ and $\tilde c_i$. From these two representations,
we can form a
new representation $T$, with column lengths
\begin{equation}
\left\{ \begin{array}{ll}
N - \tilde c_{L + 1 - i}, &  i \leq L\\
c_{i - L}, & i > L
\end{array}\right\},
\label{eq:}
\end{equation}
where $L$ is the number of boxes in the first row of the
Young tableau for $S$.  We will refer to this representation as $T =
\bar{S}R$, or as the ``composite'' representation of $R$ and $S$.
Note that this composite representation contains $L$ columns with
${\cal O}(N) $ boxes.   An example of a composite representation is
shown in Figure~\ref{f:composite}.
Computing the quadratic Casimir
(\ref{eq:casimir}) of the composite representation $T$, we find that
\begin{equation}
C_2 (T) = C_2 (R) + C_2 (S)+ \frac{2n\tilde n}{N} .
\label{eq:sumcasimir}
\end{equation}
Similarly, one can show that the dimension of the composite
representation factorizes, up to $1/N^2$ corrections (see the Appendix),
\begin{equation}
\dim T =  \dim R \dim S \left[ 1 + O({1 \over N^2}) \right],
\label{eq:dproduct}
\end{equation}
and therefore behaves, for large $N$, as
\begin{equation}
\dim T= \frac{d_R d_S N^{n +\tilde n }}{n! \tilde n!} \left[ 1 + O({1 \over
N^2}) \right].
\label{eq:leading}
\end{equation}
{}From (\ref{eq:sumcasimir}), we see that the composite representation
$\bar{S}R$ has a quadratic Casimir with leading order term $(n +\tilde n)N$.
Thus, in order to include all terms proportional to $e^{-
\frac{n\lambda A}{2}}$ in (\ref{eq:expansion}), it is necessary to
include all composite representations in the partition function sum.
We have then,
\begin{equation}
{\cal Z}  (G,\lambda A, N)  =  \sum_{n}\sum_{\tilde n}\sum_{R \in Y_n}
\sum_{S \in Y_{\tilde n}}^{} (\dim \bar{S}R)^{2 - 2G}
  e^{-\frac{\lambda A}{2 N} \left[  C_2(R)+ C_2 (S)+ \frac{2n \tilde n}{N}
\right]}.
 \label{eq:cpartition}
\label{eq:}
\end{equation}
{}From (\ref{eq:dproduct}), we see that this partition function can be
factored into a product of two copies of (\ref{eq:partfunct}), except
for the $1/N^2$ corrections from the expansion of the
dimensions, and a coupling term
$e^{- \frac{\lambda A n\tilde n}{N^2} }$.  For the torus, the factorization is
complete except for the coupling term.  In the next section, we will
prove that the partition function (\ref{eq:partfunct}) can be
rewritten as a sum over coverings of ${\cal M} $ with a fixed
orientation.  Since there are two possible orientations for an
oriented covering, we can interpret the two factors of
(\ref{eq:partfunct}) in  (\ref{eq:cpartition}) as ``chiral sectors''
of a string theory which correspond to orientation-preserving and
orientation-reversing maps of oriented 2-manifolds onto ${\cal M} $.
The coupling term $e^{- \frac{\lambda A n\tilde n}{N^2} }$ then gives a
mechanism by which these covering maps can be combined into connected
covering maps of indefinite relative
orientation.  We will return to this
interpretation of the theory in section 4.

We will now set up the specific mathematical problem which we will
address in the next section.  Ultimately, we wish to have a
geometrical description of the coefficients in the $1/N$ expansion of
the free energy (\ref{eq:expansion}).  We will find it more
convenient, however, to first work with the coefficients in the $1/N$
expansion of the partition function itself,  which are defined through
\begin{equation}
{\cal Z}  (G,\lambda A, N) = \sum_{g = -\infty}^{\infty}  \sum_{n} \sum_{i}
\eta_{g,G}^{n,i} \;
e^{- {n \lambda A\over  2}}(\lambda A)^{i} N^{2 - 2g}.
\label{eq:}
\end{equation}
We can perform a similar expansion for the partition function of a
single chiral sector,
\begin{equation}
Z (G,\lambda A, N) = \sum_{g = -\infty}^{\infty}  \sum_{n} \sum_{i}
\zeta_{g,G}^{n,i} \;
e^{- {n \lambda A\over  2}}(\lambda A)^{i} N^{2 - 2g}.
\label{eq:pchiral}
\end{equation}
It  follows from  (\ref{eq:partfunct}), that when $  2(g - 1)= 2n (G-
1)+i$, the coefficients in (\ref{eq:pchiral}) are given by
\begin{equation}
 \zeta_{g,G}^{n,i} =   \sum_{R} \left( \frac{n!}{d_R}\right)^{2G - 2}
{1 \over i!} \left( \frac{
\tilde{C}(R)}{2} \right)^{i}.
\label{eq:zeta2}
\end{equation}
The other coefficients $\zeta_{g,G}^{n,i}$ are produced from the lower
order terms in the $1/N$ expansion of the dimensions of the
representations, and the term $n^2/N$ in the quadratic Casimir
formula.  Again, in the case of the torus, there are no corrections
from dimensions, so the only other terms arise from the quadratic
Casimir.  For the purposes of the argument in the next section, we
will temporarily ignore the term $n^2/N$.  This is equivalent to
changing the gauge group to $U(N)$.  We will restore the extra term to
the quadratic Casimir when we discuss the coupling of the two chiral
sectors in section 4.

Thus, we now wish to show that the coefficients given in
(\ref{eq:zeta2}) can be interpreted in terms of covering maps.
We are interested in covering maps which have a fixed winding number
$n$, and which are only singular at points.  For now, we need only
consider singularities which arise from branch points of the form
encountered in the 2-fold covering $z \rightarrow z^2$ of the unit
disk in the complex plane.
Let us consider the set $\Sigma(G,n,i)$ of $n$-fold covers of ${\cal
M}_G$ with $i$
branch points.  If $\nu:  {\cal M}_g\rightarrow {\cal M}_G  $ is
such a covering map,  then $2(g-1) = 2n(G-1) +i$. We  include here covering
spaces which are disconnected, so that
$g$ may be negative (we define the genus of a disconnected surface
with Euler characteristic $\chi$ by $2 - 2 g = \chi$).
To every cover $\nu$, we associate a symmetry factor $| S_{\nu}|$,
which is defined to be the number of distinct homeomorphisms  $\pi$
from ${\cal M}_g $ to ${\cal M}_g $ such that $\nu \pi = \nu$.
What we shall prove is that
\begin{equation}
i! \zeta_{g,G}^{n,i} = \sum_{\nu \in \Sigma(G,n,i)}\frac{1}{| S_{\nu}|} .
\label{eq:zeta}
\end{equation}

\section{Counting Branched Covers}
\setcounter{equation}{0}

We will now proceed to prove equation (\ref{eq:zeta}) by
counting the number of branched covers of  ${\cal M}_G$ .  We first
consider the case with no branch points ($i = 0$).

For a fixed 2-manifold ${\cal M} $ of genus $G$ and of area $A$, we
choose a point $p \in {\cal M}$.
We choose a set of generators $a_{1}, b_1, a_2,b_2,\ldots$ for
$\pi_{1}({\cal M} , p)$ whose images in $H_1({\cal M})$ form a canonical
homology basis; i.e., $a_i\cdot b_j =\delta_{ij}$, $a_i \cdot a_j=b_i
\cdot b_j = 0$.  With this set of generators, the single relation
necessary to define $\pi_{1}({\cal M})$ is
\begin{equation}
a_1 b_1 a_1^{-1} b_1^{-1} a_2 b_2 a_2^{-1} b_2^{-1} \ldots
a_G b_G a_G^{-1} b_G^{-1}=1.
\label{eq:relation}
\end{equation}

Given an $n$-fold unbranched covering $\nu$ of ${\cal M}$, by choosing
a labeling  of the sheets of $\nu$ over $p$ with the
integers $I =\{1,2, \ldots, n\}$ one can construct a map from
$\pi_{1}({\cal M})$ to the permutation group $S_n$.  This map is
constructed by associating with each element $t \in \pi_{1}({\cal M})$
the permutation on $I$ which arises from transporting the labels on
sheets around the paths defined by lifting $t$ to the covering space.

{}From the definition of a covering space, this map is defined
independently of which path is chosen to represent $t$, and defines a
homomorphism  $H_{\nu}: \pi_{1}({\cal M}) \rightarrow S_n$.  In fact, it is not
hard to show that all such homomorphisms arise from some covering of
${\cal M}$ \cite{Zeischang}.  For a fixed covering $\nu$ of ${\cal M}$,
there are $n!$ possible labelings which can be used for the sheets
over $p$.  Two labelings which differ by an element $\rho$ of the
permutation group $S_n$ will give rise to homomorphisms $H$ and $H' = \rho H
\rho^{-1}$ related through conjugation by the permutation
$\rho$.  Each element of the symmetry group $S_{\nu}$ gives rise to a
permutation $\rho$ which leaves $H_{\nu}$ invariant.  Thus, the number
of distinct homomorphisms  $H: \pi_{1}({\cal M}) \rightarrow S_n$
associated with a fixed cover $\nu$ with symmetry factor $|S_{\nu}|$
is $n!/| S_{\nu}|$.  It follows that to count each cover $\nu$ with a
weight of $1/|S_{\nu}|$, it will suffice to sum over all distinct
homomorphisms $H: \pi_{1}({\cal M}) \rightarrow S_n$ with a constant
weight of $1/n!$.

Since the only defining relation of $\pi_{1}({\cal M})$ is
(\ref{eq:relation}), we can now write the weighted sum over unbranched
covers as
\begin{equation}
\sum_{\nu\in \Sigma(G,n,0)} 1/|S_{\nu}| = \sum_{s_1, t_1,  \ldots, s_G,t_G \in
S_n}^{}
\left[ \frac{1}{n!} \delta (\prod_{i=1}^{G}  s_i t_i s_i^{-1}
t_i^{-1}) \right],
\label{eq:sum}
\end{equation}
where $\delta$ is a Kronecker delta function defined on $S_n$ by
 \[
 \delta (\rho) =
 \left\{\begin{array}{ll}
 1, & \rho ={\rm identity}\\
 0, & \rho \neq {\rm  identity} .
 \end{array}\right.
 \]

Let us illustrate this for the case of the torus ($G=1$).
Here we must count the number of permutations, $s$ and $t$,  of $n$ sheets
that commute ($sts^{-1}t^{-1}=1$).  If $s$ is a permutation corresponding to
a given partition  of $n$
then there are precisely $n!/C_s $ permutations $t$ that commute with
it, where $C_s $ is the number of distinct permutations
corresponding to this partition (the order of the conjugacy class of $s$).
This is because the permutation group acts
transitively by conjugation on the set $P$ of
permutations corresponding to a fixed
partition. Thus the number of commuting elements is $n!/C_s $, and the number
number of pairs $s$ and $t$ is equal to the number of
partitions of $n$ (which we denote by $p(n)$) times $n!/C_s$   times $C_s $.
Then
we find that $\sum_{\nu\in \Sigma(1,n,0)} 1/|S_{\nu}| = p(n)$. Thus the leading
terms in the torus partition function  are given by $\sum_n  p(n)e^{- n \lambda
A} = \prod_{k=1}^\infty(1-e^{- k \lambda A})^{-1}$. This result coincides with
the evaluation in \cite{gross}.

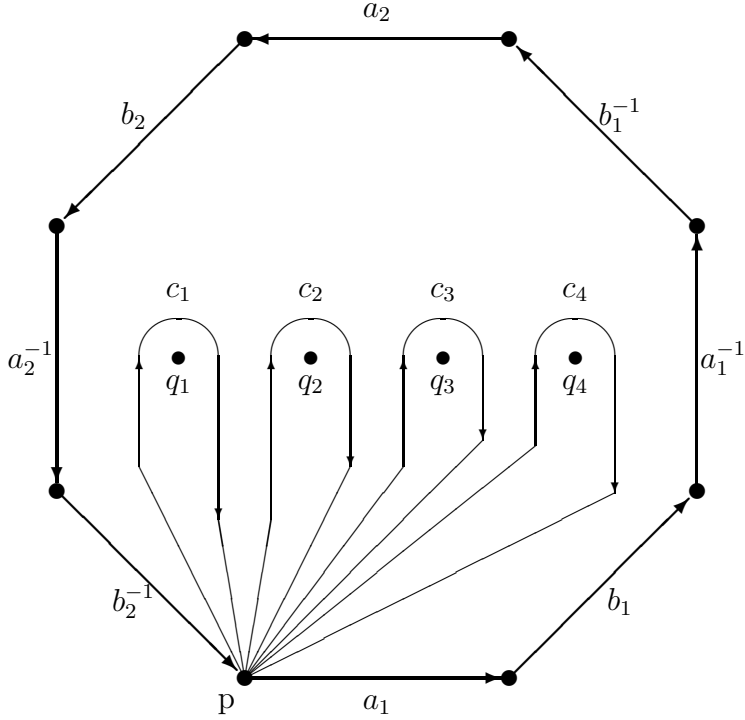
\begin{figure}[htp]
\centering
\begin{picture}(270,270)(-135,-135)
\thicklines
\put(- 50,-121){\vector(1,0){97}}
\put(50,121){\vector(-1,0){97}}
\put(121,- 50){\vector(0,1){97}}
\put(-121,50){\vector(0,-1){97}}
\put( 50,-121){\vector(1,1){68}}
\put(-50,121){\vector(-1,- 1){68}}
\put(121,50){\vector(- 1,1){68}}
\put(-121,-50){\vector(1,-1){68}}
\thinlines
\put(- 50,-121){\circle*{6}}
\put(50,121){\circle*{6}}
\put(121,- 50){\circle*{6}}
\put(-121,50){\circle*{6}}
\put( 50,-121){\circle*{6}}
\put(-50,121){\circle*{6}}
\put(121,50){\circle*{6}}
\put(-121,-50){\circle*{6}}

\put(0,-131){\makebox(0,0){$a_1$}}
\put(92,- 92){\makebox(0,0){$b_1$}}
\put(131,0){\makebox(0,0){$a_1^{-1}$}}
\put(92,92){\makebox(0,0){$b_1^{-1}$}}
\put(0,131){\makebox(0,0){$a_2$}}
\put(-92,92){\makebox(0,0){$b_2$}}
\put(-131,0){\makebox(0,0){$a_2^{-1}$}}
\put(-92,-92){\makebox(0,0){$b_2^{-1}$}}

\multiput(- 75,0)(50,0){4}{\circle*{5}}
\multiput(- 75,0)(50,0){4}{\oval(30,30)[t]}
\put(- 90,- 41){\vector(0,1){41}}
\put(- 50,- 121){\line(-1,2){40}}

\put(- 60,0){\vector(0,-1){61}}
\put(- 50,- 121){\line(-1,6){10}}

\put(- 40,- 61){\vector(0,1){61}}
\put(- 50,- 121){\line(1,6){10}}

\put(- 10,0){\vector(0,-1){41}}
\put(- 50,- 121){\line(1,2){40}}

\put(10,-41){\vector(0,1){41}}
\put(- 50,- 121){\line(3,4){60}}

\put(40, 0){\vector(0,- 1){31}}
\put(- 50,- 121){\line(1,1){90}}

\put(60, -33){\vector(0,1){33}}
\put(- 50,- 121){\line(5,4){110}}

\put(90,0){\vector(0,- 1){51}}
\put(- 50,- 121){\line(2,1){140}}

\put(- 57,- 131){\makebox(0,0){p}}

\put(- 75,- 10){\makebox(0,0){$q_1$}}
\put(- 25,- 10){\makebox(0,0){$q_2$}}
\put(25,- 10){\makebox(0,0){$q_3$}}
\put(75,- 10){\makebox(0,0){$q_4$}}

\put(- 75,25){\makebox(0,0){$c_1$}}
\put(- 25,25){\makebox(0,0){$c_2$}}
\put(25,25){\makebox(0,0){$c_3$}}
\put(75,25){\makebox(0,0){$c_4$}}
\end{picture}
\caption[x]{\footnotesize Surface with Genus $G = 2$, $i = 4$ Branch
Points}
\label{f:8gon}
\end{figure}

We shall evaluate the general sum shortly, however let us now consider the case
where there are $i\neq 0$ branch points.
We count points with branching number $j$ as  consisting of $j$ distinct branch
points which have coincided, so that for instance the map from $S^2$
to $S^2$ given by $w = z^3$ is described as having $i =4$ branch
points, although it actually has two branch points, each with
branching number 2 (0 and $\infty$).  The surface ${\cal M} $ can be
cut along the curves $a_1, b_1, \ldots$ to make a $4G$-gon with the
$i$ branch points $q_1,\ldots,q_i$.  These branch points can be chosen
in $A^i/i!$ ways, which gives rise to the extra factor of $i!$  in
(\ref{eq:zeta}).  We can construct a set of closed curves
$c_1, \ldots, c_i$ on the $4G$-gon, such that $c_j$ passes through
$p$, and is homotopic to a
loop around $q_j$, and such that the curves $c_j$ do not intersect
either one another or the curves $a_k,b_k$ except at $p$.
(For example, the case $G = 2$, $i =4$ is shown in
figure~\ref{f:8gon}.)  The curves $a_1, \ldots, a_G, b_1, \ldots, b_G,
c_1, \ldots, c_i$
form a complete set of generators for $\pi_{1}({\cal
M}\setminus\{q_1, \ldots q_i\})$, which is defined by the single
relation
\begin{equation}
c_1 c_2 \ldots c_i
a_1 b_1 a_1^{-1} b_1^{-1} a_2 b_2 a_2^{-1} b_2^{-1} \ldots
a_G b_G a_G^{-1} b_G^{-1}=1.
\label{eq:relationt}
\end{equation}
Just as in the unbranched case, given an
$n$-fold cover $\nu$ of ${\cal M} $
which is branched at the points $q_1,\ldots,q_i$, a labeling of the
sheets of $\nu$  at $p$ gives a homomorphism from $\pi_{1}({\cal
M}\setminus\{q_1, \ldots q_i\})$  to $S_n$.
The difference, however, is that the permutations $p_1, \ldots, p_i$
associated with the curves $c_j$ must be in the conjugacy class $P_n$ of
permutations which switch only two elements, since we are assuming
that all branch points have branching number 1.  We can now
generalize the expression (\ref{eq:sum}) to include branched covers by
the formula

\begin{equation}
\sum_{\nu\in \Sigma(G,n,i)} 1/|S_{\nu}| =
\sum_{p_1, \ldots, p_i \in P_n}
\sum_{s_1, t_1,  \ldots, s_G,t_G \in S_n}
\left[ \frac{1}{n!} \delta (p_1\ldots p_i \prod_{j=1}^{g}  s_j t_j s_j^{-1}
t_j^{-1}) \right].
\label{eq:branched}
\end{equation}

The sum (\ref{eq:branched}), where  the case $i=0$  reduces to our previous sum
(\ref{eq:sum}) , can be evaluated using standard results
from the representation theory of the permutation group $S_n$.  We
define the matrix associated with an element $\rho \in S_n$ in the
representation $R$ to be $D_R (\rho)$; the character is then given by
$\chi_R (\rho) = {\rm Tr}\; D_R (\rho)$.  Standard results from group
theory tell us that
\begin{equation}
\delta (\rho) = \frac{1}{n!} \sum_{R}^{} d_R \chi_R (\rho),
\label{eq:compl}
\end{equation}
\begin{equation}
\sum_{\rho \in S_n}^{}  \chi_R (\rho) D_R (\rho^{-1}) = \frac{n!}{d_R}
I_R,
\qquad{\rm and}  \qquad
\sum_{\sigma \in S_n}^{} D_R (\sigma \rho \sigma^{-1}) =
\frac{n!}{d_R}  \chi_R (\rho) I_R,
\label{eq:conjugate}
\end{equation}
where $I_R$ is the identity matrix in the representation $R$.
{}From (\ref{eq:conjugate}), it follows that
\begin{equation}
\sum_{\rho \in P_n} D_R (\rho) =
\frac{n (n - 1)}{2 d_R}  \chi_R (P) I_R,
\label{eq:conjugatet}
\end{equation}
where $\chi_R (P)$ is the character of any element of $P_n$ in the
representation $R$.
We can now use   (\ref{eq:compl})  to rewrite (\ref{eq:branched}) as
\begin{equation}
\sum_{\nu}^{}  1/|S_{\nu}| =
\sum_{p_1, \ldots, p_i}
\sum_{s_1, \ldots, t_G}
(\frac{1}{n!})^{2} \sum_{R}^{} d_R \chi_R (p_1 \ldots p_i\prod_{i}^{}
s_i t_i s_i^{-1} t_i^{-1}).
\label{eq:sum2}
\end{equation}
{}From (\ref{eq:conjugate}) and (\ref{eq:conjugatet}), we have
\begin{equation}
 \sum_{s,t \in S_n}^{} D_R(s t s^{-1} t^{-1})   =   \sum_{s,t}^{} D_R
(s t s^{-1}) D_R (t^{-1})   =  \sum_{t \in S_n}^{} \frac{n!}{d_R} \chi_R (t)
D_R (t^{-1})
I_R =  (\frac{n!}{d_R})^{2} I_R.
\end{equation}
We can
now rewrite (\ref{eq:sum2}) as
\begin{eqnarray}
\sum_{\nu\in\Sigma(G,n,i)}  1/|S_{\nu}| & = & \sum_{R}
(\frac{1}{n!})^{2} d_R
(\frac{n!}{d_R})^{2G}
\left( \frac{n (n -1) \chi_R (P)}{2d_R}  \right)^{i} {\rm Tr}\; I_R\\
 & = & \sum_{R}^{}  (\frac{n!}{d_R})^{2G - 2}
\left( \frac{n (n -1) \chi_R (P)}{2d_R}  \right)^{i}.
\end{eqnarray}
In order to prove (\ref{eq:zeta}), from (\ref{eq:zeta2}) it will
suffice to demonstrate that
\begin{equation}
\tilde{C}(R) = \frac{n (n - 1) \chi_R (P)}{d_R}.
\label{eq:casimirx}
\end{equation}
This relation can be proven as follows: clearly the operator
\begin{equation}
\tilde{P} = \sum_{p \in P_n} p
\label{eq:}
\end{equation}
is diagonal in the representation $R$, since it commutes with every
element of $P_n$ and hence with every element of $S_n$.
Assume that the representation $R$ has been defined by first
antisymmetrizing with respect to the columns (labeled by $c_1, c_2,\ldots
c_k$), of the Young tableau
associated with $R$, and then symmetrizing with respect to rows, (labeled by
$n_1, n_2,\ldots n_l$), for
some specific legal labeling of the boxes with the integers $1,
\ldots, n$
\cite{Weyl}.
Choosing a
normalized vector $v$ in the representation space of $R$, it is a
straightforward exercise to calculate
\begin{equation}
v^{T} \cdot \tilde{P} \cdot v =
\sum_{j =1}^{l}  \frac{n_j (n_j -1)}{2} -
Q_{R} \sum_{j =1}^{k}  \frac{c_j (c_j -1)}{2 Q_{R}}
= \frac{\tilde{C} (R) }{2},
\label{eq:diagonal}
\end{equation}
where
\begin{equation}
Q_{R} = \prod_{j = 1}^{l}  (n_j)!.
\label{eq:}
\end{equation}
The first term in (\ref{eq:diagonal}) is the number of pairs with
respect to which $v$ is symmetrized, and whose corresponding
permutations take $v$ to itself with eigenvalue $+1$ (these are pairs
which appear in the same row of the Young tableau ).  The second term
arises in a similar fashion from the pairs with respect to which $v$
is antisymmetrized.  This factor is slightly more complicated however,
since the vector $v$ is a sum of $Q_{R}$ separate terms, each of which
is antisymmetric with respect to a different set of $\sum_{j}c_j (c_j -
1)/2$ pairs.

Since $\tilde{P}$ is diagonal, it follows that
\begin{equation}
{\rm Tr}\; \tilde{P} =\frac{n (n - 1)}{2}  \chi (P) = \frac{\tilde{C}
(R)d_R}{2} ,
\label{eq:}
\end{equation}
from which (\ref{eq:casimirx}) follows immediately.  Thus, we have proven
(\ref{eq:zeta}).

\section{Tubes and Contracted Handles}
\setcounter{equation}{0}

We will now discuss the effects of the remaining parts of the
quadratic Casimir, and the coupling term between the two sectors; we will
show that combining these effects with (\ref{eq:zeta}) allows us to
completely interpret the coefficients $\omega_{g,G}^{n,i}$ from
(\ref{eq:expansion}) in terms of a string theory  in the case of a
toroidal target space.

To begin with, recall that in the partition function for a single
chiral sector, we have neglected the effects of the term $n^2/N$ in
the quadratic Casimir.  This term gives a multiplicative
contribution of ${e}^{ {\lambda A n^2\over 2N^2} }$ to terms in the
chiral partition function corresponding to representations with $n$
boxes in their Young tableau.  It has been pointed out by Minahan
\cite{Minahan} that these extra terms have a simple
geometric interpretation in terms of mappings where a handle of the
covering space ${\cal M}_g $ is mapped to a point in ${\cal M}_G$, or
where a pair of branch points coincide to form a ``tube'' connecting
two sheets of the cover. To see this write the extra contribution to
the exponent as ${n \lambda A \over 2N^2 } + {n(n-1) \lambda A \over 2
N^2 }$ .  The first term can be associated with the contribution of a
handle in ${\cal M}_g$ that is mapped entirely onto a single point on
the target space and which lies  on one of the $n$ sheets  of the cover. This
interpretation accounts for the factor of $n$,   for the factor of $1/N^2$
(since the
genus increases by one), the factor of $\lambda A$ (the position of
the handle) and the factor of $1/2$ (the indistinguishability of the
two ends of the handle.  The second term can be associated with
infinitesimally small tubes, connecting two sheets of the covering
space at a single point in ${\cal M}_G$ (accounting for the factor
$\lambda A$). Again this increases the genus by one (accounting for
the $1/N^2$), and the two ends of the tube can be located on any pair
of sheets of the n-sheeted cover of the target space (accounting for
the factor $n(n-1)/2$.) Since these contributions to the maps are
local they clearly exponentiate.  Note that these tubes are
essentially equivalent to combining two branch points at a single
point.  Thus, the tubes are orientation-preserving, in the sense that
moving through such a tube preserves the orientation of the covering
surface relative to the orientation of the target space.
This is consistent with our interpretation of a single chiral sector
as corresponding to covering maps with a consistent relative orientation.

The chiral partition function (\ref{eq:partfunct}) on the torus
can thus be completely understood in terms of a sum over
orientation-preserving covering maps.  We will now choose to interpret
the two copies of the chiral partition function which are coupled in
(\ref{eq:cpartition}) as arising from sums over orientation-preserving
and orientation-reversing maps.  The coupling term $-\lambda An \tilde n/N^2$,
then, can be interpreted as coupling an orientation-preserving cover
with $n$ sheets with an orientation-reversing cover with $  \tilde n$ sheets.
Just like the extra terms in the quadratic Casimir for a single
representation described above, this term is exponentiated.  We can
interpret it as describing infinitesimal orientation-reversing tubes
which connect
an arbitrary sheet of the $  \tilde n$-sheeted cover with an arbitrary sheet of
the $n$-sheeted cover.
The factor of $\lambda A$ arises
as usual from
the arbitrary location of this infinitesimal tube, and there is no
symmetry factor because the two ends of the tube are distinguished.
In addition we must introduce  a multiplicative
factor of $-1$ for each such tube.

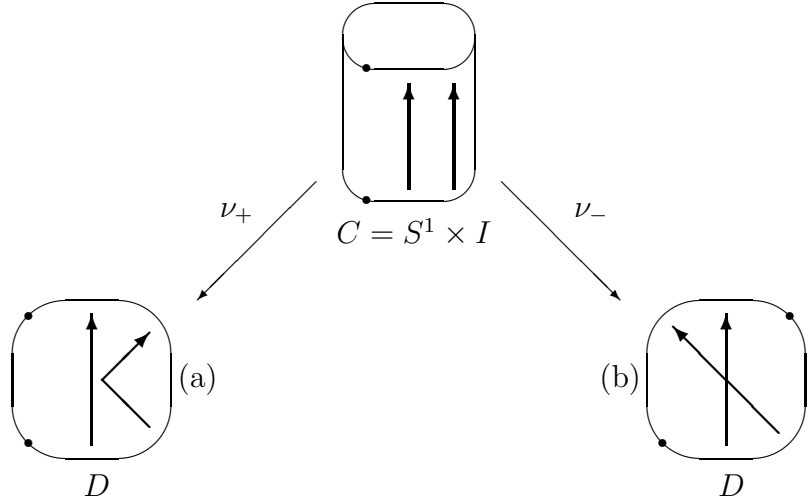
\begin{figure}[tbp]
\centering
\begin{picture}(300,200)(-170,-100)

\put(0,90){\oval(50,25)}
\put(0,40){\oval(50,25)[b]}
\put(-25,40){\line(0,1){50}}
\put( 25,40){\line(0,1){50}}
\put(0,16){\makebox(0,0){ $C = S^1\times I$}}

\put(-120,-40){\oval(60,60)}
\put(120,-40){\oval(60,60)}
\put(- 80,- 40){\makebox(0,0){(a)}}
\put(80,- 40){\makebox(0,0){(b)}}
\put(- 120,- 80){\makebox(0,0){ $D$}}
\put( 120,- 80){\makebox(0,0){ $D$}}

\put(- 35,35){\vector(-1,-1){45}}
\put(- 67,23){\makebox(0,0){ $\nu_{+}$}}
\put( 67,23){\makebox(0,0){ $\nu_{-}$}}
\put( 35,35){\vector(1,-1){45}}
\thicklines
\put(0,32){\vector(0,1){40}}
\put( 17,32){\vector(0,1){40}}
\put(- 16,28){\circle*{3}}
\put(- 16,78){\circle*{3}}
\put(-120,- 65){\vector(0,1){50}}
\put(- 144,-64){\circle*{3}}
\put(- 144,-16){\circle*{3}}

\put(- 98,- 58){\line(-1,1){18}}
\put(-116,- 40){\vector(1,1){18}}
\put(120,- 65){\vector(0,1){50}}
\put( 144,-16){\circle*{3}}
\put( 96,-64){\circle*{3}}
\put(140,- 60){\vector(-1,1){40}}

\end{picture}
\caption[x]{\footnotesize Covering Maps with Orientation-Preserving (a)
and  Reversing (b) Tubes}
\label{f:tubes}
\end{figure}

In order to clarify the distinction between orientation-preserving and
orientation-reversing tubes, we will now give a simple example of
each.  Consider the two maps from the cylinder $C = S^1\times I
=\{z,x: | z | =1, 0 \leq x \leq 1\}$ to the disk $D =\{z: | z | \leq
1\}$ given by
\begin{eqnarray}
\nu_{-} (z,x)& = &  z (1- 2x),\\
\nu_{+} (z,x) & = &  \left\{\begin{array}{ll}
z (1-2 x), x \leq \frac{1}{2} & \\
\bar{z}(2x - 1),x > \frac{1}{2}  &
\end{array}\right. .
\end{eqnarray}
These maps are both continuous $2$-fold covering maps from $C$ to $D$
which are singular only at the point 0 in $D$.  Both sheets of the
covering $\nu_{+}$ have the same relative orientation with respect to
a fixed orientation on $D$; we therefore see that the tube above the
singularity in the covering $\nu_{+}$ connects two sheets of the same
orientation, and we refer to this as an orientation-preserving tube.
Similarly, the two sheets of $\nu_{-}$ have opposite relative
orientation, so the tube in this covering is orientation-reversing.
These covering maps are illustrated in figure~\ref{f:tubes}; for
clarity we have expanded the singularity of the orientation-preserving tube.

To summarize, we can now express the entire partition function for the
$SU(N)$ gauge theory in terms of a sum over the set of
disconnected covering
maps $\Sigma_G$,
\begin{equation}
{\cal Z}(G, \lambda A, N)= \sum_{\nu \in \Sigma_G}
\frac{(-1)^{\tilde t}}{| S_{\nu}|}
e^{- \frac{n \lambda A}{2} }
\frac{(\lambda A)^{(i+ t +\tilde t+ h)}}{i!t!\tilde t ! h!}
N^{n(2- 2G)- 2(t +\tilde t +  h) - i} \left[  1
+ {\cal O}\left(  {1\over N } \right) \right],
\label{eq:total}
\end{equation}
where $n$ is the winding number of $\nu$, $i$ is the number of branch
points, $t$ ($\tilde t$) is the number of
orientation-preserving (reversing) tubes,
and $h$ is the number of handles which are mapped to
points.  Note that the ${\cal O}({1\over N }) $ corrections do
not depend on the area. For the torus there are no
correction terms  and this expansion is exact.

Now that this result is proven for the partition function, the standard
argument from quantum field theory
can be applied to prove that logarithm of the partition function corresponds to
sums over the set of connected covering
maps $\tilde{\Sigma}_G$, {\it i.e.}, we have
\begin{equation}
{\cal W}(G, \lambda A, N)= \sum_{\nu \in \tilde{\Sigma}_G}
\frac{(-1)^{\tilde t}}{| S_{\nu}|}
e^{- \frac{n \lambda A}{2} }
\frac{(\lambda A)^{(i+ t +\tilde t+ h)}}{i!t!\tilde t ! h!}
N^{n(2- 2G)- 2(t +\tilde t +  h) - i} \left[  1
+ {\cal O}\left(  {1\over N } \right) \right],
\label{eq:disconnect}
\end{equation}

\section{Plaquettes and Gluing}
\setcounter{equation}{0}

In this section we will rederive the results of section 2
from a more geometrical point of view.  A standard approach to
evaluating the gauge theory partition function (\ref{eq:partition}) on
an arbitrary manifold ${\cal M}$ is to compute the partition function
on a single plaquette (a disk with boundary $S^1$), and by gluing
together plaquettes, form a manifold with the same topology and area
as ${\cal M}$\cite{kazkos}.  This is more or less the procedure we will follow
here; we begin by showing that the partition function for a single
plaquette can be described in terms of branched covers of the
plaquette.  We then show that gluing together plaquettes combines
the separate partition functions in a way which is exactly described
by gluing together the covering spaces of the plaquettes; by gluing
plaquettes together to form the manifold ${\cal M}$, we reproduce the
results of the last section.  Although this section is essentially a
rederivation of the previous results, the development here may give
a more geometric insight into the structure of the theory.  In
particular, by using a basis of traces of the holonomy of the gauge
field around closed loops, the role of the two chiral sectors as
describing orientation-preserving and -reversing covers is made explicit.
In addition, the formalism developed here is  of use in analyzing
the theory on
manifolds with boundary, and in understanding the
effects of Wilson loops.

In this section, we will again discuss the partition function of a
single chiral sector (\ref{eq:partfunct}), and we will again drop the
term $n^2/N$ from the quadratic Casimir.  The discussion of the
previous section applies to everything done here.

Given a plaquette $\Delta$ of area $A$, if we fix the holonomy of the
gauge field around the boundary of $\Delta$ to be $U$, the partition
function of the  gauge theory on $\Delta$
is given by
\begin{equation}
{\cal Z}_{\Delta}(U) =  \sum_{R} (\dim R)\chi_{R} (U)
e^{-\frac{\lambda A}{2 N} C_2 (R)}.
\label{eq:plaquettec}
\end{equation}
This partition function is often taken as the definition of the gauge
theory on a triangulated manifold, since it is invariant under
subtriangulations, and reduces to the Yang-Mills theory in the
continuum limit \cite{migdal,witten}.

The characters $\chi_{R} (U)$ form a natural basis for the set of
class functions on the gauge group.
In order to understand (\ref{eq:plaquettec}) in terms of covering
maps, it will be useful to work with a different basis for this space.
For every partition $n = n_1+ \ldots + n_k$, there is an associated
class of elements of the symmetric group,  consisting of those
permutations with cycles of length $n_1, \ldots, n_k$.  If $\sigma$ is
such a permutation, we define
\begin{equation}
\Upsilon_{\sigma} (U) = \prod_{j = 1}^{k} ({\rm Tr}\; U^{n_j}).
\label{eq:}
\end{equation}
The set of functions $\Upsilon_{\sigma}(U)$ also form a complete
basis for the set of class functions on the gauge group.  These
functions are related to the characters by the relations
\begin{eqnarray}
\chi_R (U) & = & \sum_{\sigma\in S_n}\frac{\chi_{R}(\sigma)}{n!}
\Upsilon_{\sigma} (U),  \label{eq:class} \\
\Upsilon_{\sigma}(U) & = &  \sum_{R} \chi_{R} (\sigma)
\chi_{R} (U).
\end{eqnarray}

Just as we did for the partition function (\ref{eq:partition}) in
section 3, we can expand ${\cal Z}_{\Delta} $ as a power series
in $1/N$.  Once again, the partition function consists of two coupled
chiral sectors.  For the moment, we will consider a single chiral
sector, where the partition function is given by
\begin{equation}
Z_{\Delta}(U) =  \sum_{n} \sum_{R\in Y_n} (\dim R)\chi_{R} (U)
e^{-\frac{\lambda A}{2 N} C_2 (R)}.
\label{eq:plaquette}
\end{equation}
For a single sector, the leading order terms in $N$, for fixed values
of $n$ and $A$, are given by
\begin{equation}
Z_{\Delta} (U) =
\sum_{n}e^{-\frac{ n \lambda A}{2} }
\sum_{i} \frac{(\lambda A)^i}{i!}
\sum_{R} \left[\frac{d_R}{n!}
\left( - \frac{\tilde{C}(R)}{2}  \right)^i \chi_{R} (U)
N^{n - i} + {\cal O}( N^{n - i - 1} ) \right].
\label{eq:expand}
\end{equation}
This can be rewritten in terms of the class functions
$\Upsilon_{\sigma}$, as
\begin{equation}
Z_{\Delta} (U) =
\sum_{n}e^{-\frac{ n \lambda A}{2} }
\sum_{i}\frac{(\lambda A)^i}{i!}
\sum_{\sigma}\left[\frac{(-1)^i}{C_{\sigma}}  \phi^{\sigma}_{n,i}
\Upsilon_{\sigma} (U) N^{n - i}+ {\cal O}( N^{n - i - 1} ) \right],
\label{eq:expand2}
\end{equation}
where $C_{\sigma}$ is the number of permutations in the conjugacy
class of $\sigma$.  Plugging (\ref{eq:class}) into (\ref{eq:expand}),
we find
\begin{equation}
\phi^{\sigma}_{n,i} =\sum_{R}\frac{d_{R} C_{\sigma}}{n!^2}
\left(\frac{ \tilde{C}(R)}{2}  \right)^i\chi_{R} (\sigma).
\label{eq:phi}
\end{equation}

We will now show that $\phi^{\sigma}_{n,i}$ is given by a sum over
covers similar to that for $\zeta_{g,G}^{n,i}$.  Specifically, we
define $\Sigma_{\sigma}(n,i)$ to be the set of $n$-fold covers of
$\Delta$ with $i$ branch points and with the additional property that the
boundary of the covering space is a disjoint union of $k$
copies of $S^1$ which cover the boundary of $\Delta$ $n_1, \ldots, n_k$
times -- as above, $n_j$ are the sizes of the cycles of the
permutation $\sigma$.  We will show that
\begin{equation}
\phi^{\sigma}_{n,i} = \sum_{\nu \in \Sigma_{\sigma}(n,i)}\frac{1}{|S_{\nu}|}.
\label{eq:}
\end{equation}

{}From an argument analogous to that leading to (\ref{eq:sum}) and
(\ref{eq:branched}), we can write the sum over covers as
\begin{eqnarray}
\sum_{\nu\in \Sigma_{\sigma}(n,i)} 1/|S_{\nu}|  & = &
\sum_{p_1, \ldots, p_i \in P_n}
\left[ \frac{C_{\sigma}}{n!} \delta (p_1\ldots p_i \sigma) \right]\\
 & = & \sum_{R}\frac{d_{R} C_{\sigma}}{n!^2}
\left( \frac{ \tilde{C}(R)}{2}  \right)^i\chi_{R} (\sigma),
\end{eqnarray}
so the assertion is proven.

As an example, consider the term
\begin{equation}
\phi^{P}_{n,1}= \frac{1}{2 (n - 2)!},
\label{eq:example}
\end{equation}
where $P\in P_n$ is a permutation consisting of a single pair
exchange.   For any value of $n$, the only covering space of $\Delta$
with a single branch
point has a boundary which consists of $n - 2$ single covers of the
boundary of  $\Delta$, and one double cover.  Thus, $P$ is in the only
conjugacy class with $\phi^{\sigma}_{n,1}\neq 0$. The symmetry factor
of the unique cover is exactly $2 (n - 2)!$, which accounts for the
denominator in (\ref{eq:example}).

To summarize what we have shown so far in this section, we can write
the highest order terms in the chiral partition function
(\ref{eq:plaquette})
 in terms of a sum
over all coverings $\nu$ of $\Delta$,
\begin{equation}
Z_{\Delta} (U)= \sum_{\nu}\left[  \frac{(-1)^i}{| S_{\nu}|}
e^{- \frac{n \lambda A}{2}}
\frac{(\lambda A)^i}{i!} \prod_{j} ({\rm Tr}\;  \hat{U}_j) N^{n - i}
+ {\cal O}(N^{n - i - 1}) \right],
\label{eq:single}
\end{equation}
where $n$ is the winding number of $\nu$, $i$ is the number of branch
points, and $\hat{U}_j$ are the holonomies of the pullback of the gauge
field to the covering space.  We will now show that this
formula continues to hold when we glue plaquettes together, so that for an
arbitrary orientable manifold ${\cal M} $ of genus $G$, area $A$, and with $l$
boundary components, the highest order terms in the chiral partition
function are given by a sum over all
coverings of ${\cal M} $,
\begin{equation}
Z_{\Delta} (U_1, \ldots, U_l)= \sum_{\nu}\left[
\frac{(- 1)^i}{| S_{\nu}|}
e^{- \frac{n \lambda A}{2} }
\frac{(\lambda A)^i}{i!} \prod_{j} ({\rm Tr}\;  \hat{U}_j)
N^{n(2- 2G- l) - i}
+ {\cal O}(N^{n(2 - 2G - l) - i - 1}) \right],
\label{eq:general}
\end{equation}
where again $\hat{U}_j$ are the holonomies of the pullback of the
gauge field around the boundaries of the covering space.
Note that this formula could also have been derived directly by arguments
similar to those above; it is perhaps more instructive, however, to
derive this result through the following argument.

We will prove (\ref{eq:general}) by induction on the number of
internal edges in a triangulation of ${\cal M} $.
(By a triangulation, we mean a decomposition into plaquettes which
are joined along edges -- the plaquettes need not be triangular;
an internal edge is
an edge which is not on the boundary).  The result
(\ref{eq:single}) is the case where there is a single plaquette,
with 0 internal edges.  What we now need to show is that when we add a
new internal edge, either by gluing together two manifolds ${\cal M} $
and ${\cal N} $ along a common edge, or by gluing together two edges
of a single manifold ${\cal M} $, the relation (\ref{eq:general})
continues to hold.  The approach we use to glue together placquettes
is similar to methods which have been widely used in the context of
$U(N)$ lattice gauge theories in the large  $N$ limit\cite{kazkos} .

In order to glue together the partition function along an edge, we use
the integral
\begin{equation}
\int dU \mixten{U}{i_1}{j_1}\ldots \mixten{U}{i_n}{j_n}
\mixten{U}{\dagger k_1}{l_1} \ldots\mixten{U}{\dagger k_n}{l_n}  =
\frac{1}{N^n} \left( \sum_{\sigma \in S_n}
\mixten{\delta}{i_1}{l_{\sigma_1}}  \mixten{\delta}{k_{\sigma_1}}{j_1}
\ldots
\mixten{\delta}{i_n}{l_{\sigma_n}}  \mixten{\delta}{k_{\sigma_n}}{j_n}
\right)+ {\cal O}  (N^{-n- 1}),
\label{eq:gluing}
\end{equation}
where $\sigma = (\sigma_1, \ldots ,\sigma_n)$ ranges over all
permutations of the integers $1, \ldots, n$ (we have normalized $\int
dU =1).$ We glue along an edge by integrating over the integral  of the
gauge field along the edge.
Formula (\ref{eq:gluing}) is essentially the statement that to highest
order in $N$, the
products of traces of $U$ combine according to the Wick expansion
familiar from matrix model theories.  This formula has a simple geometric
interpretation, which is that when two plaquettes are glued
together along a common boundary, their covers are glued together in
all possible ways
consistent with their boundary structure.  The induction step follows
immediately from this integral, as can be seen from the following
arguments:

If we take two manifolds, ${\cal M} $
and ${\cal N} $, the set of $n$-fold coverings of the glued manifold ${\cal
M}\cup {\cal N}  $ is given by taking all $n$-fold covers of
${\cal M} $ and gluing them to all $n$-fold covers of ${\cal N} $ in
all possible ways along the common edge.  The winding number is fixed,
and the genus and the number of branch points are additive.  The
holonomies around the boundaries of the new covering space are
formed by contracting the holonomies around the old boundaries in exactly the
way specified by (\ref{eq:gluing}).  The number of edges on the glued
manifold is one less than the sum of the numbers of edges on ${\cal M}
$ and ${\cal N} $, so the factor of $N^{- n}$ in (\ref{eq:gluing})
fixes the exponent of $N$ in (\ref{eq:general}) correctly.  Finally, a
short combinatorial
argument shows that the symmetry factors combine in the correct
fashion to give the symmetry factor of the glued cover.

In a similar way, one can go through the details of gluing a single
manifold ${\cal M} $ to itself along an internal edge.
There are two possible forms that such a gluing might take.
If the two copies of the internal edge
which are being glued together are in the same boundary component, the
gluing has the net effect of adding one to the number of edges.  The
factor of $N^{- n}$ from (\ref{eq:gluing}) correctly fixes the
exponent of $N$ in (\ref{eq:general}).  In the other case, the two
edges are in different boundary components.  In this case, the genus
is increased by one, and the number of edges decreases by one.  Again,
the power of $N$ is fixed correctly.
Finally, one can contract  an edge to a point.  This decreases the
number of edges by one and contributes a factor of $N^{n}$ to the
partition function.

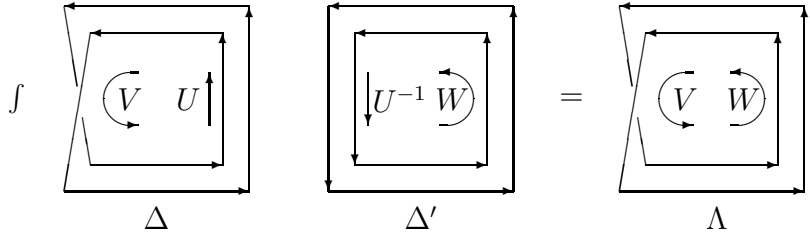
\begin{figure}[tbp]
\centering
\begin{picture}(200,100)( - 135,- 50)
\multiput(- 170,- 35)(210,0){2}{\vector(1,0){ 70}}
\multiput(- 160,- 25)(210,0){2}{\vector(1,0){ 50}}
\multiput(- 110,- 25)(210,0){2}{\vector(0,1){50}}
\multiput(- 110,25)(210,0){2}{\vector(-1,0){ 50}}
\multiput(- 160,25)(210,0){2}{\line(-1,-6){10}}
\multiput(- 100,- 35)(210,0){2}{\vector(0,1){70}}
\multiput(- 100,35)(210,0){2}{\vector(-1,0){ 70}}
\multiput(- 170,35)(210,0){2}{\line(1,-6){5}}
\multiput(- 160,-25)(210,0){2}{\line(-1,6){3}}

\put(- 70,- 35){\vector(1,0){ 70}}
\put(- 60,- 25){\vector(1,0){ 50}}
\put(- 10,- 25){\vector(0,1){50}}
\put(- 10,25){\vector(-1,0){ 50}}
\put(- 60,25){\vector(0, -1){50}}
\put(0,- 35){\vector(0,1){70}}
\put(0,35){\vector(-1,0){ 70}}
\put(- 70,35){\vector(0,-1){70}}

\put(75,0){\makebox(0,- 90){ $ \Lambda$}}
\put(- 35,0){\makebox(0,- 90){$\Delta'$}}
\put(-135,0){\makebox(0,- 90){$\Delta$}}
\put(- 190,0){\makebox(0,0){ $\int$}}
\put(20,0){\makebox(0,0){ $=$}}
\multiput(-145,0)(210,0){2}{\makebox(0,0){$V$}}
\multiput(- 145,0)(210,0){2}{\oval(20,20)[l]}
\multiput(-145,- 10)(210,0){2}{\vector(1,0){3}}
\multiput(-145,10)(210,0){2}{\line(1,0){3}}
\put(- 125,0){\makebox(0,0){ $U$}}
\put(- 115,-  10){\vector(0,1){20}}
\put(-45,0){\makebox(0,0){ $U^{-1}$}}
\put(- 55,  10){\vector(0,-1){20}}
\multiput(- 25,0)(110,0){2}{\makebox(0,0){ $W$}}
\multiput(- 25,0)(110,0){2}{\oval(20,20)[r]}
\multiput(- 25,10)(110,0){2}{\vector(-1,0){3}}
\multiput(- 25,- 10)(110,0){2}{\line(-1,0){3}}

\end{picture}
\caption[x]{\footnotesize Geometric Interpretation of
$\int ({\rm Tr}\; VUVU) ({\rm Tr}\; U^{-1} W)^2$}
\label{f:trace}
\end{figure}

As an example, let us consider
the gluing together of two plaquettes $\Delta$ and $\Delta'$
along a common edge to form
a new plaquette $\Lambda$.  We will take the areas of the plaquettes
$\Delta$ and $\Delta'$ to be $A$ and $A'$.
The partition function on $\Lambda$ can be written as
\begin{equation}
Z_{\Lambda}  (VW) = \int d U
Z_{\Delta}(VU)Z_{\Delta'}(U^{-1} W),
\label{eq:combine}
\end{equation}
where $VW,VU,$ and $U^{-1} W$ are the holonomies of the gauge field
around the boundaries of $\Lambda, \Delta,$ and $\Delta'$ respectively.
{}From (\ref{eq:expand2}) and (\ref{eq:phi}), we can write the first few
terms in the partition functions for $\Delta$ and $\Delta'$.
\begin{eqnarray}
Z_{\Delta}(VU)  & = &
e^{-\frac{ \lambda A}{2} } \left[({\rm Tr}\; VU) N + \ldots\right]\nonumber \\
& & +e^{- \lambda A}\left[
\frac{1}{2} ({\rm Tr}\; VU)^2 N^2 -
\frac{\lambda A}{2} ({\rm Tr}\; VUVU) N + \ldots
 \right] \nonumber \\
& & +e^{-\frac{3\lambda A}{2}} \left[
\frac{1}{6} ({\rm Tr}\; VU)^3 N^3 + \ldots \right] + \ldots
\\
Z_{\Delta'}(U^{-1} W)  & = &
e^{-\frac{\lambda A}{2} } \left[({\rm Tr}\; U^{-1} W) N +
\ldots\right]\nonumber \\
& & +e^{- \lambda A}\left[
\frac{1}{2} ({\rm Tr}\; U^{-1} W)^2 N^2 -
\frac{\lambda A}{2} ({\rm Tr}\; U^{-1} WU^{-1} W) N + \ldots
 \right] \nonumber \\
& & +e^{-\frac{ 3\lambda A}{2}} \left[
\frac{1}{6} ({\rm Tr}\; U^{-1} W)^3 N^3 + \ldots \right] + \ldots
\end{eqnarray}
Plugging these expressions into (\ref{eq:combine}), we can compute the
integral term by term;
\begin{eqnarray}
\int d U e^{-\frac{ \lambda A}{2} } e^{-\frac{ \lambda A'}{2} }
({\rm Tr}\; VU)({\rm Tr}\; U^{-1} W)N^2 & = &
e^{-\frac{\lambda (A+ A')}{2} }  ({\rm Tr}\; VW) N,
\\
  \int d U e^{- \lambda A} e^{- \lambda A'}
\frac{1}{4} ({\rm Tr}\; VU)^2({\rm Tr}\; U^{-1} W)^2 N^4   &= &
 e^{- \lambda (A+ A')}\frac{1}{2}  ({\rm Tr}\; VW)^2 N^2 + {\cal O}(N) ,
\end{eqnarray}
\begin{eqnarray}
  \int d U   e^{- \lambda A}  e^{- \lambda A'}
\frac{1}{2} N^3 \left[ \frac{\lambda A}{2}
({\rm Tr}\; VUVU)({\rm Tr}\; U^{-1} W)^2  +
 \frac{\lambda A'}{2}
({\rm Tr}\; VU)^2 ({\rm Tr}\; U^{-1} WU^{-1} W) \right]
\label{eq:double}\\
 =  e^{- \lambda (A+ A')} \frac{\lambda (A + A')}{2}
({\rm Tr}\; VWVW) N + {\cal O}  (1).
\label{eq:double2}
\end{eqnarray}
Combining these results, we get
\begin{eqnarray}
Z_{\Delta}(VW)  & = &
e^{-\frac{ \lambda (A + A')}{2} } \left[({\rm Tr}\; VW) N +
\ldots\right]\nonumber \\
& & +e^{- \lambda (A + A')}\left[
\frac{1}{2} ({\rm Tr}\; VW)^2 N^2 -
\frac{\lambda (A + A')}{2} ({\rm Tr}\; VWVW) N + \ldots
 \right] \nonumber \\
& & +e^{-\frac{3\lambda (A + A')}{2}} \left[
\frac{1}{6} ({\rm Tr}\; VW)^3 N^3 + \ldots \right] + \ldots, \label{eq:lambda}
\end{eqnarray}
which is exactly what we expect, since the partition function is
invariant under subtriangulation.  As a particular example, we will
now describe the interpretation of (\ref{eq:double2}) from the
geometric point of view as a gluing of covering spaces.  This term
as it appears in the partition function (\ref{eq:lambda}) is
associated with covers of $\Lambda$ with winding number 2 and 1 branch
point.  The two terms in (\ref{eq:double}) correspond to the two
ways such a cover can be constructed by gluing covers of $\Delta$ and
$\Delta'$; let us take the first term, which arises from gluing
together a cover of $\Delta$ with one branch point with a cover of
$\Delta'$ with no branch points.  This gluing is illustrated in figure
{}~\ref{f:trace}.  There is a single 2-fold cover of $\Delta$ with a single
fixed branch point.  The branch point can be anywhere on $\Delta$,
which accounts for the factor of $\lambda A$.  This cover has a symmetry
factor of $2$.  There
is also a single cover of $\Delta'$ with no branch points, also with a
symmetry factor of 2.  These two covers can be joined in exactly two
possible ways along the common edge.  Both of these joinings, however,
give topologically equivalent covers of $\Lambda$.  The factor of 2
fixes the correct symmetry factor of this single cover, which is $2$.
Thus, the first term in (\ref{eq:double}) corresponds to a cover of
$\Lambda$ with a single branch point, which lies in $\Delta$.  The
other term in (\ref{eq:double}) corresponds similarly to covers of
$\Lambda$ with a branch point lying in $\Delta'$.

To review the results of this section: we have shown by induction that
(\ref{eq:general}) describes the
highest order terms in the chiral partition function for an arbitrary
manifold ${\cal M} $ with boundary as a sum over all coverings of the
manifold ${\cal M} $.  In particular, we have reproduced the result of
section 3 by a more geometric argument.
Note that the fact that (\ref{eq:general}) is exact for the torus is
not explained by this argument, in which we have only retained highest
order terms in $N$.  It is interesting to note that the only other
target space for which (\ref{eq:general}) has no lower order
corrections is the annulus (the genus 0 surface with two boundary
components.)

We can now discuss what happens when we return the extra terms to the
quadratic Casimir and couple the two chiral sectors.  The effects of
the term $n^2/N$ and the coupling term $\lambda A n \tilde n/N^2$ are the same
as was discussed in the previous section.  In the case of a single
plaquette, or of any manifold with boundary, however, there is now an
extra coupling, due to the fact that
\begin{equation}
\chi_{\bar{S}R} (U) \neq \chi_{S} (U) \chi_{R} (U).
\label{eq:}
\end{equation}
In fact, $\chi_{\bar{S}R} (U) $ is the character of the only
irreducible representation occurring in the tensor product
representation $\bar{S}\otimes R$ which has a quadratic Casimir with
leading term $(n + \tilde n)N$.  The effect of this coupling is precisely to
cancel any folds which might occur by contracting  a factor  of $U$
with a factor of $U^{\dagger}$ from boundaries of covering sheets with
opposite orientations.  This can be seen as follows:

We have shown
that the representations $R$ with $n$ boxes can be associated with
linear combinations of $n$th order invariant polynomials in $U$, which
are then associated with contours around a covering of a placquette
with a fixed orientation.  If one had a contour around the boundary in
the opposite direction, one would have a factor of $U^{\dagger}$.  In
fact, the representations $\bar{S}$ are exactly the complex conjugates
of the representations $S$, and therefore their characters are linear
combinations of polynomials of order $  \tilde n$ in $U^{\dagger}$.  This gives
a natural understanding of why one chiral sector involves covers with
one orientation, and the other chiral sector involved covers with the
opposite orientation.
To return to the question of folds, the effect of the extra coupling
implicit in the appearance of $\chi_{\bar{S}R}$ in the partition
function is to subtract all terms where any factors of $U$ and
$U^{\dagger}$ from the same plaquette are contracted.  From the point
of view developed in this section, this is exactly the suppression of
folds which is responsible for allowing the theory to be described in
terms of maps without folds.

Finally, let us note that the results
achieved here for manifolds with boundary are also of use in
discussing the partition function when Wilson loops are inserted on
the target space manifold.  This subject will be described in a later paper.

\section{Non-Orientable Target Spaces}
\setcounter{equation}{0}

So far we have restricted our attention to orientable target spaces. However it
is easy to extend the discussion of QCD and of the string theory to the
non-orientable case. The partition function for QCD on a general non-orientable
manifold has been derived by Witten \cite{witten}. The main difference is that
only {\em self-conjugate} representations ($R=\bar R $) contribute. Thus, for
example, the   partition function for a manifold consisting of a genus $G$
surface to which $K$ copies of the Klein bottle are  attached is,
\begin{equation}
Z_{G,K}= \sum_{R=\bar R} \left( \dim R \right)^{2-2G-2K}e^{-\frac{\lambda A}{2
N}C_2(R)} .
\label{eq:klein}
\end{equation}
The representations that survive in the large $N$ limit are now the composite
representations discussed in section 2, except that $S=R$ and $n=\tilde n$.
Because of this we see that only even winding numbers
occur. The Casimir operator of these self-conjugate representations
simplifies,
\begin{equation}
C_2(\bar R R) = 2( nN + \tilde C( R) ) .
\label{eq:cascoj}
\end{equation}
Therefore, in the $1/N$ expansion the terms that we associated with collapsed
handles and tubes cancel completely, due to the $(-1)$ associated with
orientation reversing tubes. Thus, for example, the partition function of the
Kein bottle, ($G=0, K=1$), is given by
\begin{equation}
Z_{0,1}= \sum_{R=\bar R} e^{ - n\lambda A - {\lambda A \tilde C( R) \over N}} ,
\label{eq:kleint}
\end{equation}
 which can be interpreted as precisely a sum over even branched covers of the
Klein bottle.

{}From the point of view of the orientable string theory we have
developed here, the fact that $n = \tilde{n}$ for non-orientable surfaces
has a simple geometric interpretation  in terms of covering maps.
Any cover, connected or disconnected, of a non-orientable surface
by an orientable surface, must always have an equal number of sheets
with each of the two possible relative orientations.  This is because
if we consider the permutation on sheets associated with a curve
around which the orientation of the target space is reversed, this
permutation must take sheets with one relative orientation to sheets
with the other relative orientation.  Thus, each cycle for such a
permutation must contain an equal number of sheets with each
orientation.  It is therefore clear, that the partition function of a
single chiral sector over a non-orientable surface vanishes, and that
only terms with $n = \tilde{n}$ contribute to the complete theory.

\section{Conclusions}
\setcounter{equation}{0}

We will now briefly review the status of the general program of
interpreting 2-dimensional gauge theories as string theories.
We have shown here that the coefficients $\omega_{g,G}^{n,i}$ in the
asymptotic expansion of the partition function for the $SU(N)$ gauge
theory on a genus $G$ surface have a
simple interpretation as a sum over maps from a genus $g$ surface to a
genus $G$ surface, when $2(g-1) = 2n(G-1) +i$.  When the target space is a
torus ($G = 1$), we can understand all coefficients $\omega_{g,G}^{n,i}$
in terms of such maps, so we have a complete understanding of the
geometric structure of this partition function.
Since this is the
physical case of interest, namely a flat target manifold, we can claim with
confidence that QCD$_2$ is equivalent to a theory of maps of a two dimensional
internal space onto  the target space--{\it i.e.} a string theory.

When the genus of ${\cal M} $ is not $1$, extra terms arise in the
partition function from the lower order terms in the $1/N$ expansion
of the dimensions,  corrections to Equation  (\ref{eq:dimension}).  These terms
give rise to extra contributions to the
coefficients $\omega_{g,G}^{n,i}$ when
\begin{equation}
2 (g - 1) > 2n (G - 1) + i.
\label{eq:}
\end{equation}
We do not yet have a geometric understanding of these terms which can
be related to a string theory picture. Since they do not occur on the torus
they must be related to the global properties of the target space.

For those terms which we do understand, in particular for the case of
the torus, the natural next step is to attempt to write down a string
theory whose partition function is equal to the free energy of the
gauge theory.  As discussed above, the natural conclusion of this paper is that
the string action is something like the Nambu action plus a term that totally
suppresses folds but otherwise does not contribute, say by contributing
infinite (zero) action  when the map has (does not have) folds. What  is this
term? One  possibility is to simply introduce a constraint into the functional
integral that eliminates folds. This can be done by adding to the action a
term,
\begin{equation}
S_{\rm fold} = \int d^2\xi \sqrt{g} \lambda\left( n-1 \right),
\label{eq:fold}
\end{equation}
where $\lambda$ is a Lagrange multiplier field and $n$ is the normal to the
embedded surface, $n={\rm sign}\left[ \det{ \left(  {\partial {x^\mu}\over
\partial \xi^\alpha}\right)} \right]$. In two dimensions the normal is a
scalar, taking values $\pm 1 $, and the discontinuities occur at the folds.
However this is not a very elegant term and  it does not generalize to higher
dimensions. Another possibility is that there exist extra fermionic fields on
the surface, which have zero modes for the folded maps and therefore eliminate
them from the sum. We would have to add in addition corresponding bosonic
fields that would cancel the contribution of the fermions for allowed maps.
These fields might give extra corrections for target spaces with $G\neq 1$,
thus accounting for the extra terms in the $1/N$ expansion of the dimensions.
They might also explain the factor of $(-1)$ that occurs for
orientation-reversing tubes.

        The next step in understanding the string picture of QCD$_2$ is to
incorporate fermions.  First, one should calculate the correlation functions of
Wilson loops. These can be calculated for arbitrary loops. However, for
complicated loops, on arbitrary manifolds, the calculation is very involved.
One needs to expand, in powers of $1/N$, not just the dimensions and Casimirs
of general  representations of $SU(N)$, but also their 6-j symbols. Also, these
correlation functions depend not just on the total area and genus of the
manifold but also on the areas inclosed by all non self-intersecting
portions of the loop. Real quarks are even more complicated. In principle they
can be treated once one knows how to handle arbitrary Wilson loops, with any
number of self intersections.

Finally one should try to construct a string theory for QCD in higher
dimensions, including four. This will no longer be so simple to evaluate. Two
dimensions is clearly a very degenerate case both for gauge theories, since
there are no glueball states -- there being no transverse dimensions, and for
string theory, due to the simple nature of maps and the apparent suppression of
folds. In higher dimensions there will be a full spectrum of glueballs.
Correspondingly we expect that the string theory will be more complex.
Presumably this is because the  term in the string action that suppresses
extrinsic curvature will now contribute to the weight of the maps. It is only
for two dimensional maps that  the extrinsic curvature is either zero or
infinity. However we are confident that a string representation exists.
In QCD we expect the physics to be continuous as we vary the dimension between
two and four.  As for string theory normally one expects trouble for $d>2$ due
to the tachyonic nature of the center of mass degree of freedom. However, for
the QCD string this mode does not appear for $D=2$, as it does for the
Liouville string, and therefore we have no reason to expect a tachyon to appear
as we go above two dimensions. The
fold-preventing  term in  the string action    keeps the string rigid,
preventing the  proliferation of thin tubes which presumably correspond to the
tachyonic mode of the string,  and thus eliminates this instability!
Also, in this two dimensional string theory there is, as in QCD$_2$, no sign of
the dilaton field, whose spatial dependence  is responsible for the breaking of
Lorenz invariance in the Liouville two-dimensional string theory. We might hope
that its generalization to higher dimensions will yield a Lorenz invariant
theory, with no gravitons or gauge mesons.

\setcounter{section}{0}
\startappendix
\section{ Dimensions and Casimirs for Large N}
\setcounter{equation}{0}

In this appendix we will analyse the dimensions and Casimirs of the
representations of $SU(N)$ that survive in the large $N$ limit.

As discussed in section 2 a   representation $R$,
associated with a Young tableau containing  rows of length $ {n}_1, \ldots,
{n}_l$ has a  quadratic Casimir
given by
\begin{equation}
C_2 (R) =   N\sum n_i + \sum_in_i(n_i-(2i-1)) - \frac{n^2}{N},
\label{eq:}
\end{equation}
The only representations which can survive as $N\to \infty$ are those which
either have a finite number of total boxes, or which have a finite number of
columns of length of order $N$. These are precisely the composite
representations, $T=\bar S R$,  considered in section 2. It is straightforward
to show that their quadratic Casimir is given by (\ref{eq:sumcasimir}).

The dimension of a representation of $SU(N)$ whose Young tableaux has
rows of length
$(n_1,n_2,\ldots n_k)$ is given by Weyl's formula,
 \begin{equation}
\dim R  = {\prod_{1\leq i<j \leq N}(h_i-h_j)\over \prod_{1\leq i<j\leq N}( i-
j)}   ; \qquad {\rm where} \,\, \,\, h_i=N+n_i-1 .
\label{eq:Weyl}
\end{equation}
As shown in \cite{gross} one can separate out of this formula the dimension
$d_R$ of the representation of the symmetric group of $n\equiv
\sum_{i=1}^{k}{n_i}$ objects, corresponding to the partition $ \left[n_1\geq
n_2\geq  \dots \geq n_k\right]$,
 \begin{equation}
  d_R= d_{\left[ n_1, n_2, \dots n_k\right]}=
n! {\prod_{1\leq i<j \leq k}(h_i-h_j)\over \prod_{1\leq i<j\leq k}( i- j)} ,
\label{eq:red}
\end{equation}
to derive
\begin{equation}\dim R =  {d_R \over n!}
 \prod_{i=1}^{k}{{(N+n_i-i)!}\over (N-i)!}
\label{eq:sym}
\end{equation}
This formula is well suited for a ${1\over N}$ expansion when the total number
of boxes, $n$, is fixed as $N\to \infty$. Note that
$
{(N+n_i-i)! \over (N-i)!}= N^{n_i}\prod_{j=1}^{n_i}{\left( 1+{j-i \over N}
\right)} ,$  where the index $j$ ($i$) in this formula runs over the columns
(rows) of the tableau. Thus we can write the dimension as,
\begin{equation}
   \dim R =  {d_R N^n \over n!} \prod_{v}   \left(  1+{\Delta_v \over N}
\right)  ,
 \label{eq:expans}
\end{equation}
where the product rums over all the cells of the tableau  and $ \Delta_v$ is
defined for each cell to be the column index minus the row index.

Now consider the dimensions of the composite representations $T=\bar S R$. It
is a straightforward algebraic exercise to separate the expression
(\ref{eq:Weyl})  into  a product over the rows of $R$ and of $ S$
respectively times a cross-term. The result is that
\begin{eqnarray}\dim T &= & \dim R  \dim S  \, Q[R,S] \\
Q[R,S] & \equiv &\prod_{i,j} { \left(  N+1 -i-j \right)        \left(   N+1
-i-j +n_i+\tilde n_j \right)    \over  \left(  N+1 -i-j +n_i\right) \left(  N+1
-i-j +\tilde n_j  \right)},
\label{eq:corection}
\end{eqnarray}
where the product is over the rows of $R$ ($S$) of length $n_i$($\tilde n_j$).
Note that whenever the representation $R$ or $S$ is trivial, then $Q=1$.
Furthermore, since both $n=\sum_i   n_i$ and $\tilde n= \sum_j \tilde n_j$ are
finite, $Q\to 1 $ as $N\to \infty$. In fact
\begin{equation}
Q[R,S]  \sim  1 - {n \tilde n \over N^2} + {n \tilde C(S) + \tilde n \tilde
C(R)
\over N^3 } + O \left(  {1 \over N^4}\right) .
\label{}
\end{equation}

\vskip .5truein
{\Large{\bf Acknowledgements}}

We would like to thank Orlando Alvarez    for helpful discussions.

\end{document}